# Confined monolayer Ag as a large gap 2D semiconductor and its momentum resolved excited states

Woojoo Lee[1,3], Yuanxi Wang[2,6], Wei Qin[1,3], Hyunsue Kim[1,3], Mengke Liu[1,3], T. Nathan Nunley[1,3], Bin Fang[3], Rinu Maniyara[4], Chengye Dong[2], Joshua A. Robinson[2,4], Vincent Crespi[2,5], Xiaoqin Li[1,3], Allan H. MacDonald[1,3] and Chih-Kang Shih[1,3, †]

[1] *Department of Physics, The University of Texas at Austin, Austin, Texas 78712, USA*
[2] *Two-Dimensional Crystal Consortium and Materials Research Institute, The Pennsylvania State University, University Park, PA, 16802, USA*
[3] *Center for Dynamics and Control of Materials, The University of Texas at Austin, Austin, Texas 78712, USA*
[4] *Materials Science and Engineering, The Pennsylvania State University, University Park, PA, 16802, USA*
[5] *Department of Physics, The Pennsylvania State University, University Park, PA, 16802, USA*
[6] *Department of Physics, University of North Texas, Denton, TX, 76203, USA*

[†] *Corresponding author: shih@physics.utexas.edu*

**2D materials have intriguing quantum phenomena that are distinctively different from their bulk counterparts. Recently, epitaxially synthesized wafer-scale 2D metals, composed of elemental atoms, are attracting attention not only for their potential applications but also for exotic quantum effects such as superconductivity. By mapping momentum-resolved electronic states using time-resolved and angle-resolved photoemission spectroscopy (ARPES), we reveal that monolayer Ag confined between bilayer graphene and SiC is a large gap (> 1 eV) 2D semiconductor, consistent with GW-corrected density functional theory. The measured valence band dispersion matches the DFT-GW quasiparticle band. However, the conduction band dispersion shows an anomalously large effective mass of 2.4 $m_0$. Possible mechanisms for this large enhancement in the "apparent mass" are discussed.**

The capability to synthesize materials with atomic layer precision has enabled many scientific discoveries and advanced technologies, exemplified by the discovery of graphene and its derivatives [1–4]. Epitaxial metal layers have also inspired discoveries of several novel quantum effects in the ultra-thin regime [5–8] although their investigations are often limited to ultra-high-vacuum environments unless the surface is properly protected and stabilized by a capping layer[9]. Recently, a novel intercalation method was developed in which confined metal layers can be realized between a SiC substrate and a bilayer graphene (BL-Gr) capping layer which prevents oxidation of the confined epitaxial metals [10,11]. More intriguingly, such a confined metal layer may harbor novel electronic properties [12,13]. For example, density functional theory (DFT) predicted that silver- and indium-monolayer (MLs) are narrow gap semiconductors due to their hybridization with the underlying SiC [14]. Indeed, recent ARPES investigations of intercalated Ag monolayer (Ag-ML) revealed the existence of a valence band maximum (VBM) below the Fermi level [15]. However, the lack of information on the electronic structure above the Fermi level has impeded this new platform to be utilized in science and technology. One of the reasons for detaining its investigation is that the conventional alkali doping method does not work effectively for graphene-based systems [16]. The Schottky contact between the semiconducting Ag-ML and the metallic graphene, makes it difficult to tune the Fermi level above the conduction band minimum (CBM) using alkali doping, thus preventing the excited states from being accessed using static-ARPES.

Here we report ARPES and time-resolved ARPES (trARPES) investigations of the electronic structures of both occupied and unoccupied states in the confined Ag-ML between a SiC substrate and BL-Gr. We find that the confined Ag-ML is indeed an artificial 2D semiconductor with a surprisingly large bandgap (~ 1 eV). Theory predicts this bandgap correctly when quasiparticle effects were accounted for (without the inclusion of graphene layer) using GW calculations on top of mean-field DFT results. The valence band dispersion is found to be consistent between the experiment and DFT-GW calculations. However, the conduction band dispersions differ substantially. The experimental effective mass is found to be a factor of 2.6 larger than the theoretical prediction. Moreover, the energy at the $\bar{M}_{Ag}$ point is revealed to be only 0.14 eV higher than the conduction band minimum (CBM) at the Γ point, in stark contrast to a predicted value of 0.6 eV in DFT-GW calculations. These unusual features are attributed to the results of excitonic effect by precluding many-body interactions between the highly n-doped BL-Gr and Ag-ML electrons.

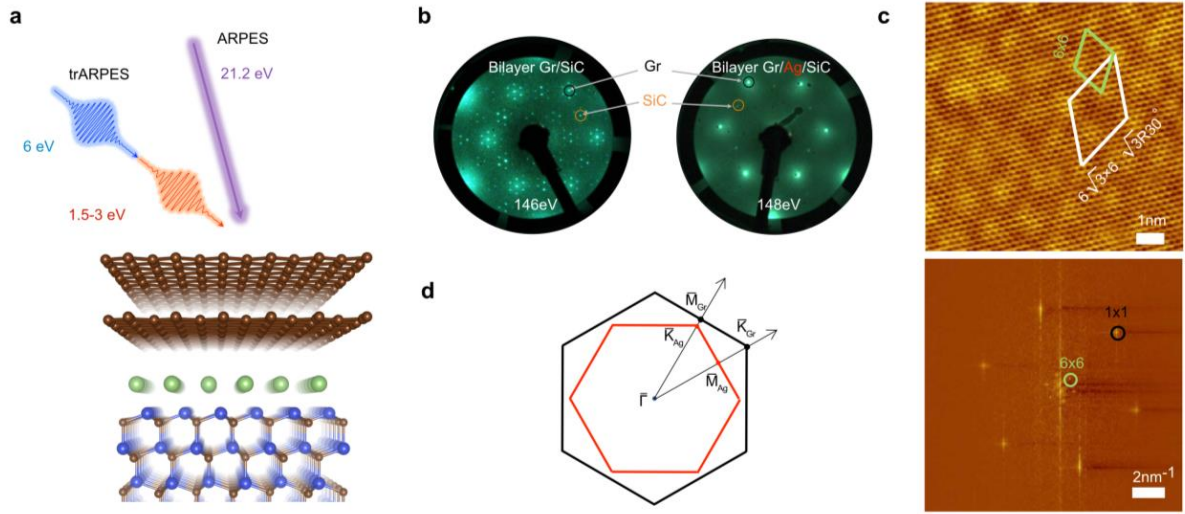

*Figure 1 Monolayer Ag atomic structure and surface characterization.* *(a) Experimental scheme and atomic structure of intercalated Ag between bilayer graphene and SiC. (b) Low energy electron diffraction (LEED) images before and after Ag intercalation. (c) STM topography (top) and its real space fast Fourier transform (FFT) image (bottom). (d) Brillouin zones of monolayer Ag (red) and bilayer graphene (black).*

The atomic structure of the confined metal monolayer and the experimental setup are illustrated in Fig. 1a. At the thermodynamic ground state, the Ag-ML assumes the lattice constant of SiC with surface Ag atoms projecting vertically onto the second topmost layer of C sites in the SiC substrate; the next most stable Ag registry (Ag projecting onto the *topmost* C sites) is higher in energy by +8 meV per Ag. Both configurations are realistic and considered in theory discussions later. The BL-Gr protects the confined epitaxial Ag from oxidation, allowing sample transport under ambient conditions. The confined growth of metal monolayers using epitaxial graphene as a template has been described previously [10]. The same procedure is used in this study to intercalate Ag atoms between the BL-Gr and SiC substrate. All samples are cleaned in the ultra-high-vacuum (UHV) chamber (pressure $< 1\times10^{-10}$ mbar) by overnight annealing at 220 ℃. A thoroughly cleaned sample exhibits a sharp graphene 1×1 LEED pattern co-existing with another pattern consistent with the SiC lattice constant (Fig. 1b). The commonly observed sharp $6\sqrt{3} \times 6\sqrt{3}R30°$ patterns for pristine BL-Gr on SiC become a fuzzy ring, suggesting a decoupling of BL-Gr and SiC due to Ag intercalation. STM studies, however, still exhibit a quasi 6×6 superstructure superimposed on the graphene 1×1 lattice (Fig. 1c), consistent with the Ag-ML following the SiC lattice constant[15]. Based on this structural information, the surface Brillouin zones (BZs) for BL-Gr and Ag-ML are determined in Fig. 1d.

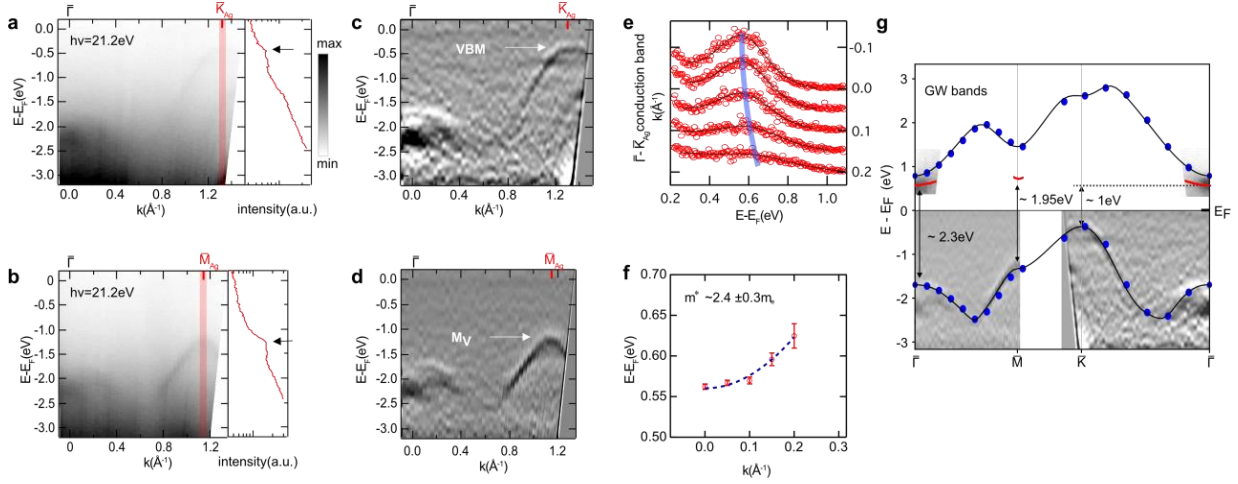

*Figure 2 Occupied and unoccupied electronic structures of monolayer Ag. (a, b) Occupied band dispersions along high symmetry directions. (c, d) Second derivative images of (a) and (b) respectively. (e) Unoccupied structure of the monolayer Ag along the $\bar{\Gamma} - \bar{K}_{Ag}$ direction. (f) The effective mass was determined from EDC fittings from (e). (g) DFT-GW calculated band dispersion superimposed with acquired experimental images. The experimental conduction band dispersion near the $\bar{\Gamma}$ point (up to $k_\parallel = 0.2\ \text{Å}^{-1}$) is marked by a red curve. The energy location at the $\bar{M}_{Ag}$ point is marked by a red bar.*

The occupied band dispersions of Ag-ML are obtained along two major directions, $\bar{\Gamma} - \bar{K}_{Ag}$ and $\bar{\Gamma} - \bar{M}_{Ag}$. The static-ARPES data using 21.2 eV photon energy is shown in Fig. 2a and 2b respectively, and their second-derivative counterparts are shown in Fig. 2c and 2d. The valence band maximum (VBM) is identified at $\bar{K}_{Ag}$ with a binding energy of ~ 0.45 eV below $E_F$. At the saddle point $\bar{M}_{Ag}$, a local maximum is identified as $\bar{M}_V \sim -1.25\ eV$. Near the $\bar{\Gamma}$ point, the SiC band structure is also observed with SiC$_{VBM}$ located around –1.8eV. The same band dispersion acquired with different photon energy (40.8eV) can be found in Fig. S1. We note that our VB mapping result is very similar to that reported recently for Ag-ML confined between ML-Gr and SiC [15], except for the absolute energy relative to the Fermi level.

Next, we reveal the unoccupied band dispersion near $\bar{\Gamma}$ using trARPES with 3 eV photons (Fig. 2e). The CBM is identified at ~ 0.56 eV above $E_F$, and its peak position does not show probe power dependence, indicating the absence of the space charge effect (Fig. S2). Moreover, the possibility that this state is induced by imaging potential is ruled out by performing pump photon energy-dependent measurements (Fig. S3). The effective mass, $2.4 \pm 0.3\ m_0$, of the conduction band is determined by mapping the dispersion from $k_\parallel = 0$ to $k_\parallel = 0.2\ \text{Å}^{-1}$ (Fig. 2f). The acquired occupied/unoccupied band dispersions reveal an indirect gap of 1.01 eV, a value significantly larger than the earlier DFT prediction of 0.2 eV [14].

To better capture this large bandgap feature, we carried out GW corrections to the DFT calculation (see details in Methods) and superimposed the results on the experimental observation as shown in Fig. 2g. To make the DFT-GW calculations computationally feasible, the graphene cap is excluded, which simplifies the computation cell to a minimal SiC surface unit cell. The calculation predicts a bandgap value of ~ 1.15 eV, consistent with the experimental observation. Moreover, the valence band dispersion from the DFT-GW calculations agrees well with the experimental results. The difference (0.15eV) between the experimental and the calculated bandgap sizes may be attributed to dielectric/Coulomb screening from graphene layers, which was not explicitly included in the GW calculations. Such a renormalization effect on graphitic substrates[17–21] has been studied intensively investigated recently.

By contrast, the calculated conduction band dispersion clearly differs from the experimental observation. The GW quasiparticle band dispersion near CBM yields an effective mass of 0.94 $m_0$, a factor of 2.6 lighter than the experimental estimate. Moreover, as discussed further below, the GW calculation predicts that the $\bar{M}_{Ag}$ point gap size is ~ 2.63eV after considering the -0.15eV error, thus a strong light absorption is expected at the same size of photon energy. However, the absorption was observed at 1.95eV.

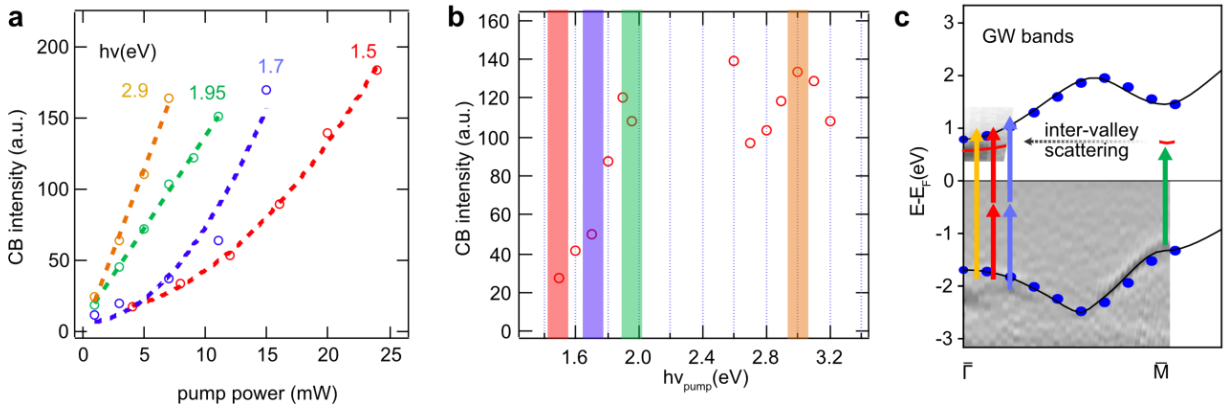

*Figure 3 Below-bandgap absorption at $\bar{M}_{Ag}$ point. (a) Pump power dependence of $\bar{\Gamma}$ point CB intensity at four pump photon energies. Linear(1.95 and 2.9eV) and quadratic (1.5 and 1.7eV) dispersions reflect one-photon absorption (1PA) and two-photon absorption (2PA) processes. (b) CB intensity versus pump photon energy at fixed 4mW pump power ($85\mu J/cm^2$). (c) Schematic of 1PA and 2PA process at $\bar{M}_{Ag}$ and $\bar{\Gamma}$ points. The smaller photon energies, $h\nu_{pump}$<1.95eV, excite the electrons to $\bar{\Gamma}$ point CB via 2PA. On the other hand, at the higher photon energy, $h\nu_{pump}$=1.95eV, the system absorbs the photons via 1PA by forming the quasi-particle bound states (excitons) which are immediately scattered to $\bar{\Gamma}$, resulting in the hot exciton dispersion.*

To gain further insight, we performed trARPES as a function of the pump photon energy. Fig. 3a shows the photoelectron intensity near the CBM as a function of pump power for four different pump photon energies. The quadratic dependence on the pump power clearly shows two-photon absorption (2PA) processes for $h\nu = 1.5 \text{ and } 1.7 \ eV$. On the other hand, for $h\nu =$ 1.95 and 2.9 $eV$, only the one-photon absorption event is involved as revealed by the linear dependence. Fig. 3(b) shows the photoelectron intensity near the CBM vs. pump energy at a constant incident pump fluence of ~ 85μJ/cm². The plot shows that as the $h\nu_{pump}$ drops below 1.95 eV, the excitation efficiency to the conduction band quickly decreases, consistent with the existence of a threshold of $h\nu \approx 1.95\ eV$ for a transition from 2PA to 1PA process.

As the direct gap at the $\bar{\Gamma}$ point is ~ 2.35 eV, it is not surprising that excitations to the CB using $h\nu = 1.5$ and $1.7\ eV$ require two-photon absorption (and for that matter, one photon absorption for $h\nu = 2.9\ eV$) (see Fig. 3c). What is surprising is that only one-photon absorption is involved for $h\nu = 1.95\ eV$ which is smaller than the direct gap at the $\bar{\Gamma}$ and $\bar{M}_{Ag}$ points. This unexpected observation leads us to consider two possible origins of the observed large effective mass at $\bar{\Gamma}$-point; a band renormalization by an interaction between plasmons on BL-Gr and electrons on Ag-ML, and an exciton-dressed band dispersion.

We first characterize the high electron density in the BL-Gr embodying collective modes of electrons (plasmon) and discuss its role on the Ag-ML's electronic system. Shown in Fig. 4a are $E$ vs. $k$ mappings for a series of k-space segments perpendicular to the $\bar{\Gamma} - \bar{K}_{Gr}$ direction. The zoomed-in image of the band mapping at $\bar{K}_{Gr}$ is shown in Fig. 4b. One can see a gap opening at the Dirac point (DP) due to the lattice symmetry breaking along the c-axis creating a vertical electric field [22,23]. The splitting between the coupled Dirac bands in BL-Gr is observable with two Fermi wavevectors, $k_{F,1} = 0.11 \pm 0.01$ Å$^{-1}$ and $k_{F,2} = 0.03 \pm 0.01$ Å$^{-1}$. The energy separation for the two Dirac bands above the DP is ~ 0.36 eV. The Fermi velocity $v_F$, fitted at $k_{F,1}$, is $0.55 \times 10^6\ cm/s$, about half of the Dirac velocity of the unperturbed Dirac cone in single-layer graphene. The Dirac electron density, $n_s$, is related to the Fermi wavevector by $n_s = k_F^2/\pi$, from which we calculate an electron density of ~ $4 \times 10^{13}/cm^2$. The BL-Gr electronic structure is consistent with previously reported results for the same density [16].

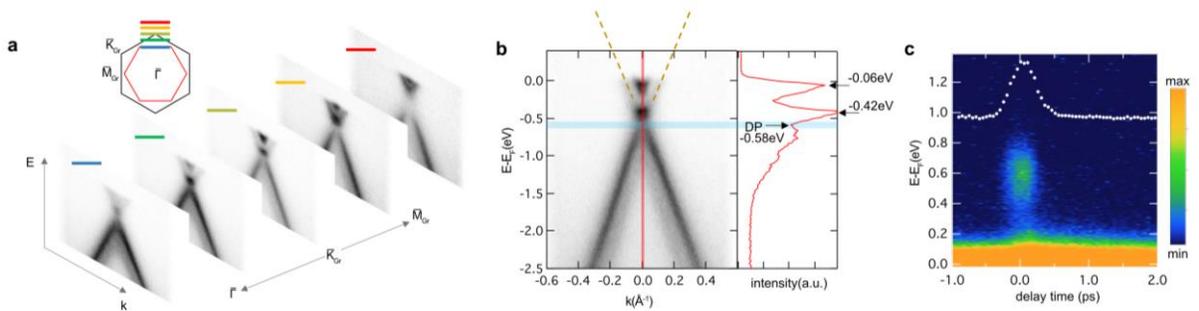

*Figure 4 High electron density in the bilayer graphene and an ultrafast decay of the Ag monolayer's CB state.* (a) Dirac cone dispersions along the $\bar{\Gamma} - \bar{K}_{Gr}$ direction. (b) Zoomed-in Dirac cone at $\bar{K}_{Gr}$ point. (c)

*Photoelectron intensity versus energy and time delay at the $\bar{\Gamma}$ point. White dots indicate the CBM intensity distribution.*

Given the high electron density in the BL-Gr, plasmons can be generated [24,25] and interact with electrons [26] in the Ag-ML. The electron-plasmon interactions are expected to renormalize electron effective mass analogous to electron-phonon interactions [27]. For quantitative analysis, we model and solve a Hamiltonian to get an interaction-induced self-energy and a renormalized mass of electrons in the Ag-ML. The detailed derivation can be found in Supplementary note 1. Despite the existence of finite mass enhancement, the result shows that it is less than 2% of the free electron mass. We attribute this inefficient enhancement to the much faster dynamics of plasmons than Ag electrons as represented in their dispersive band (Supplementary Fig. S4). Within the time scale of the Ag electron dynamics, the contributions from the plasmon modes are averaged to negligible value. This is in analogy to the electron-electron interaction-dressed electron effective mass. Although the Coulomb interaction is strong, its effect on electron effective mass is very weak and can be ignored. Thus, we conclude that the mass enhancement driven by electron-plasmon interaction cannot provide a satisfactory explanation for our experimental results. On the other hand, the ultrafast decay time of the excited state implies that the electron-plasmon coupling still plays an important role as an efficient energy relaxation channel (Fig. 4c).

We next discuss an excitonic effect on the acquired band dispersion. The exciton, a quasi-particle commonly observed in 2D semiconductors, consists of an electron-hole pair attracted via Coulomb interaction. This two-particle bound state can participate in the photoemission process by leaving the hole behind the escaped electron. The escaped electron states are expected to have a downward-parabolic dispersion in energy-momentum (E-k) space as governed by energy conservation law [28]. On the other hand, at finite temperature, the ensemble of thermally excited hot exciton states modifies the parabolic dispersion to be opened upward, mimicking a single-electron dispersion but with a heavier effective mass, $m_e + m_h$ [28]. To this end, we presume the exciton-dressed band dispersion as an origin of the experimentally determined large effective mass, $2.4 \pm 0.3\ m_0$.

Considering the exciton-dressed band dispersion, it is intriguing that the one-photon absorption, at photon energy smaller than the direct gap at the $\bar{\Gamma}$ and $\bar{M}_{Ag}$ points, led to populate the CB states. According to the DFT-GW calculations, $\bar{M}_{Ag}$ point is a saddle point in the E-k dispersion for both CB and VB, resulting in a diverging density of states. Thus the 1PA threshold of 1.95eV can be interpreted as the excitonic transition. Given the $\bar{M}_{Ag}$ point gap size of ~2.63eV, it implies an exciton binding energy of ~ 0.68eV, which is consistent with binding energies observed in various 2D semiconductors [29]. Once the excitons are formed, it is expected to be scattered transiently to the $\bar{\Gamma}$ point exciton state via inter-valley exciton scattering by leaving electrons at $\bar{\Gamma}$ valley and holes at $\bar{K}_{Ag}$ valley, i.e. momentum-forbidden dark excitons. The inter-valley exciton scattering was observed directly in monolayer WSe$_2$ and MoS$_2$ by photoemission experiments [30,31]. We expect that the same effect is manifested here. Unfortunately, in the current case, the CB state's lifetime is much shorter than our time resolution, precluding us from observing the dynamical behavior of excitons as reported for WSe$_2$ and MoS$_2$ monolayers [30,31].

In summary, we combined ARPES and trARPES to probe momentum-resolved quasi-particle dispersions for both the equilibrium and excited states and revealed that a confined Ag monolayer between SiC and BL-Gr is an artificial 2D semiconductor with an indirect band gap of 1 eV. We also showed that the equilibrium quasi-particle dispersion in the valence band dispersion is well-captured by the DFT-GW calculation. However, the excited state dispersion in the conduction band is dramatically different, exemplified by an anomalously large effective mass (a factor of 2.6 enhancement). We attribute this massive enhancement to the consequence of exciton-dressed band dispersion by precluding another scenario, mass enhancement induced by electron-plasmon interactions, which only contribute to the rapid decay time.

**Methods**

**Sample preparation**

Epitaxial Graphene (EG) is synthesized via sublimation of silicon from the (0001) plane of semi-insulating 6H-SiC (II-VI Inc.) at 1,800 °C, 700 Torr pressure, and 500 sccm Ar flow, for 30 min. EG is then plasma treated using a Tepla M4L plasma etch tool, using 150 sccm of O2 and 50 sccm of He, under a pressure of 500 mTorr and power of 50W for 60 s. Two-dimensional silver intercalation was performed using Thermo scientific Lindberg/Blue M Mini-Mite tube furnace fitted with a 1-inch outer diameter quartz tube. A custom-made alumina crucible from Robocasting Enterprises was used to hold 1×1 cm EG/SiC substrates. The EG grown face is placed downwards inside the crucible. Then, 75 mg of silver powder (Sigma Aldrich, 99.999%) was placed in the crucible directly beneath the EG/SiC substrate. The crucible with EG/SiC and the silver powder was then loaded into the tube furnace and evacuated to ~5 mTorr. The tube was then pressurized to 500 Torr with Ar. The furnace was then heated to 950°C with a ramp rate of 50° min$^{-1}$ and an Ar flow of 50 sccm. The furnace was held at the growth temperature for 1 hour, then cooled to room temperature.

# Experimental details

ARPES/ 6eV-trARPES measurements

A Scienta R3000 hemisphere analyzer was used to collect the photoemission spectra for both static-ARPES and 6eV-trARPES measurements. The static ARPES measurements were carried out with a Helium lamp using He I-$\alpha$ (21.2eV) and He II-$\alpha$ (40.8eV). During the helium lamp measurements, the pressure was better than $6 \times 10^{-10}\ Torr$. The 6eV-trARPES measurements were performed using probe pulse (206nm) and pump pulse (826~387nm). The pressure was maintained below $8 \times 10^{-11}\ Torr$. The fundamental 1030nm pulses from a Carbide laser of Light conversion were used for the 5$^{th}$ harmonic generator (HIRO) and Orpheus-HP optical parametric amplifier (OPA) to generate the pump and probe pulses at a 100 kHz repetition rate. All measurements were performed at room temperature. The pump-probe cross-correlation width was estimated as 400 fs (FWHM) via the fastest photoemission dynamics. The incident fluence was ~85μJ/cm$^2$.

# Computational details

DFT-GW calculations were based on the Ag/SiC structure described in the main text. Additional DFT calculations were performed for structures with Ag projecting onto the topmost C sites and onto the topmost Si sites, where M point energies were 0.47 and 0.80 eV relative to the CBM respectively, neither smaller than that of the main Ag/SiC structure we considered 0.45 eV. This excludes the possibility that the small M point energy could be due to a thermodynamically competing Ag surface phase. In all DFT-GW calculations, we employ an 18×18×1 k-point grid and a truncated Coulomb interaction[32]. We employ the extrapolar technique[33] to achieve convergence for all quasiparticle band energies within 0.05 eV. Extrapolar energy of 2.0 Ha was extracted from systematic convergence studies, allowing accelerated convergence at a plane-wave energy cutoff of 10 Ha and 100 empty bands for the evaluation of the dielectric matrix and self energies. The calculated static dielectric matrix was extended to finite frequencies using the Godby–Needs-generalized plasmon-pole model[34]. All calculations are performed using the ABINIT code[35].

# Acknowledgment

This work was primarily supported by the National Science Foundation through the Center for Dynamics and Control of Materials: an NSF MRSEC under Cooperative Agreement No. DMR-1720595. Other supports include NSF Grant Nos. DMR-1808751, and the Welch Foundation F-1672. Support for synthesis comes from The Penn State Center for Nanoscale Science (NSF Grant DMR-2011839) and the Penn State 2DCC-MIP (NSF DMR-1539916).

**Contributions**

W.L., J.A.R., and C.K.S. conceived the experiment. W.L. carried out trARPES/ARPES measurements. Y.W. and V.C. performed DFT-GW calculations. R.M., C.D., and J.A.R. prepared the sample. M. L. performed STM measurements. W.L. created a platform for 6eV trARPES and UHV investigation with the participation of H.K., T.N.N., and B.F. under the joint supervision of C.K.S. and X.L.. A.H.M. and W.Q. performed theoretical model calculations. W.L., C.K.S., and A.H.M analyzed the data. W.L. and C.K.S. wrote the manuscript with substantial contributions from all the authors.

**Data availability**

The data that support the findings of this study are available from the corresponding author upon reasonable request.

# Confined monolayer Ag as a large gap 2D semiconductor and its momentum resolved excited states

Woojoo Lee[1,3], Yuanxi Wang[2,6], Wei Qin[1,3], Hyunsue Kim[1,3], Mengke Liu[1,3], T. Nathan Nunley[1,3], Bin Fang[3], Rinu Maniyara[4], Chengye Dong[2], Joshua A. Robinson[2,4], Vincent Crespi[2,5], Xiaoqin Li[1,3], Allan H. MacDonald[1,3] and Chih-Kang Shih[1,3, †]

[1] *Department of Physics, The University of Texas at Austin, Austin, Texas 78712, USA*
[2] *Two-Dimensional Crystal Consortium and Materials Research Institute, The Pennsylvania State University, University Park, PA, 16802, USA*
[3] *Center for Dynamics and Control of Materials, The University of Texas at Austin, Austin, Texas 78712, USA*
[4] *Materials Science and Engineering, The Pennsylvania State University, University Park, PA, 16802, USA*
[5] *Department of Physics, The Pennsylvania State University, University Park, PA, 16802, USA*
[6] *Department of Physics, University of North Texas, Denton, TX, 76203, USA*

[†] *Corresponding author: shih@physics.utexas.edu*

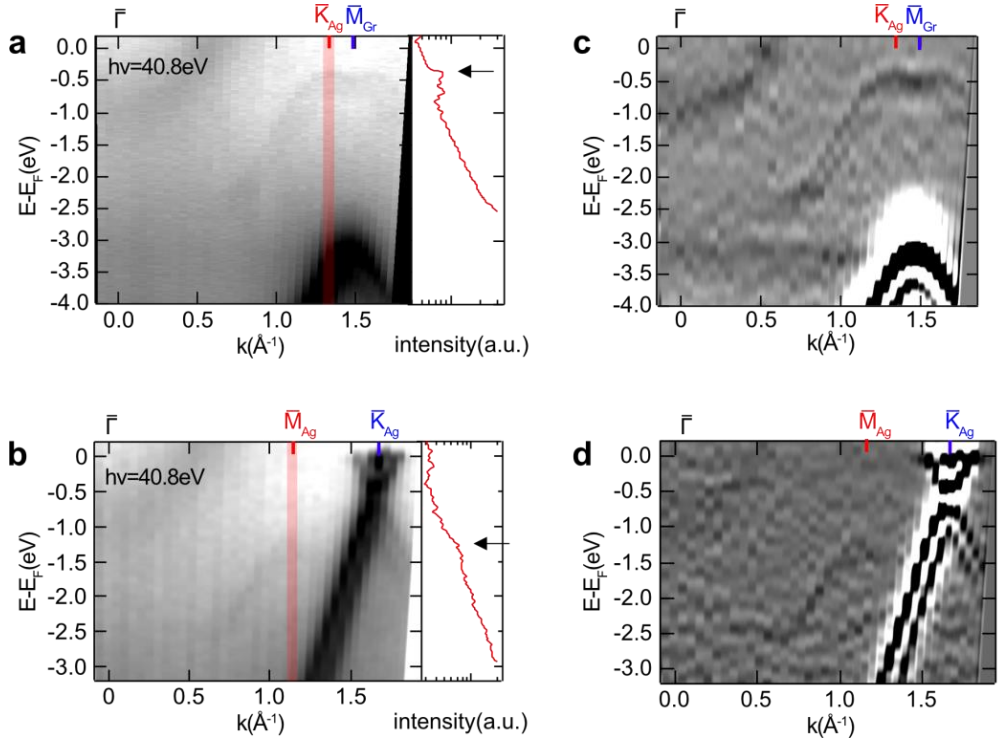

**Figure S1 Occupied electronic band structure of 2D metal (Gr/Ag/SiC) acquired with 40.8eV photon energy.** (a, b) Band dispersions were obtained by static-ARPES along $\bar{\Gamma} - \bar{K}_{Ag}$ and $\bar{\Gamma} - \bar{M}_{Ag}$ respectively. (c, d) Second derivative images of (a) and (b).

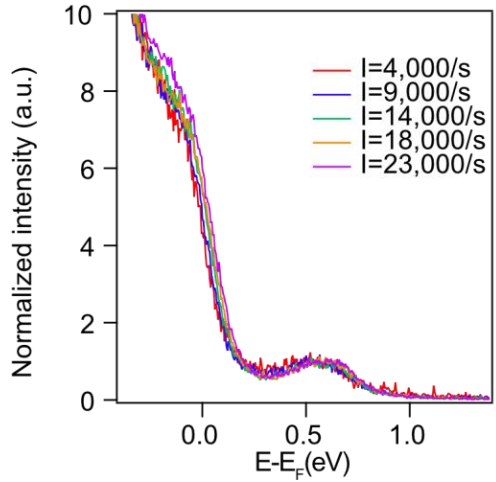

**Figure S2 6eV probe pulse power dependence at CBM.** The intensity stands for photoelectron counts per second obtained by the electron analyzer. Below I=23,000/s, no space charge effect was observed. All data of this work was taken with the probe power below I=9,000/s.

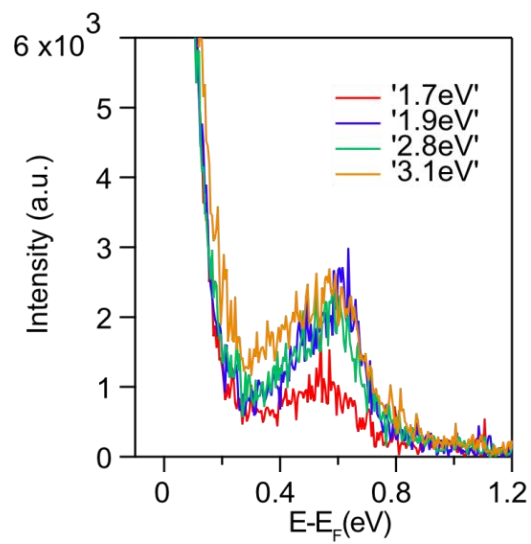

**Figure S3 Pump photon energy dependence at CBM.** The fact that the CBM peak position does not change on the different pump photon energy excludes the possibility of observing image potential states which can be excited by a 6eV pulse and detected by a pump pulse. In this case, the peak position changes depending on the pump photon energies.

**Supplementary note 1: Electron-plasmon interaction-induced mass renormalization**

In this section, we consider the mass renormalization effect on the Ag conduction band due to the coupling with plasmon modes in nearby BL-Gr. The interaction between electrons in Ag and plasmon in BL-Gr can be formulated as the following second-quantization form[S1]

$$H_{ep} = \sum_{q} M(\mathbf{q})(a_q + a^{\dagger}_{-q})c^{\dagger}_{k+q}c_k, \tag{S1}$$

where the coupling matrix $M(\mathbf{q}) = \sqrt{\hbar\omega_q v_q/(2A)}e^{-qd}$, $A$ denotes the area of the sample, $a_q$ and $c_k$ are the annihilation operators of a plasmon in BL-Gr and electron in Ag, respectively. The Coulomb potential $v_q = 2\pi e^2/\kappa q$, where the dielectric constant $\kappa \sim 4.5$ is approximated by half value of the static dielectric constant of SiC substrate [S2]. The screening effect caused by the metallic BL-Gr is captured by a factor of $e^{-qd}$ with $d \approx 0.65$ nm the vertical separating distance between monolayer Ag and BL-Gr, as depicted schematically in Fig. 1(a) in the main text. Within two-dimensional (2D) electron gas approximation, the plasmon frequency

$$\omega_q = \sqrt{n_e q^2 v_q/m^*}, \tag{S2}$$

where $n_e = 4 \times 10^{13}$ cm$^{-2}$ is the electron density in BL-Gr and $m^*$ denotes the electron effective mass. The Hamiltonian for plasmon takes the following form

$$H_{pl} = \sum_{q} \hbar\omega_q\, a^{\dagger}_q a_q. \tag{S3}$$

The effective mass of electron in BL-Gr was extracted from the temperature dependence of Shubnikov–de Haas oscillations at low carrier densities[S3]. Extrapolation of those data to electron density $n_e = 4 \times 10^{13}$ cm$^{-2}$ gives rise to $m^* \sim 0.25 m_0$, where $m_0$ is the mass of the free electron. Fig. S4(a) shows the energy dispersions plasmon for several values of $m^*$, and the small-$q$ parts are consistent in order of magnitude with earlier results obtained from random-phase calculations[S4]. In the present study, we only consider acoustic plasmon mode because the coupling between the inter-band optical plasmon mode and electron in monolayer Ag is strongly suppressed due to the layer polarization of charge in BL-Gr.

In close analogy to electron-phonon interaction, the electron-plasmon interaction-induced self-energy of the electron in monolayer Ag takes the following form

$$\Sigma(i\omega_n, \boldsymbol{k}) = \sum_q |M(\boldsymbol{q})|^2 \left[ \frac{1 + N_q - f(\epsilon_{k+q})}{i\omega_n - \epsilon_{k+q} - \hbar\omega_q} + \frac{N_q + f(\epsilon_{k+q})}{i\omega_n - \epsilon_{k+q} + \hbar\omega_q} \right], \quad (S4)$$

where $\omega_n = (2n+1)\pi/\beta$ is the fermionic Matsubara frequency with $\beta = 1/k_B T$, $N_q = 1/(e^{\beta\hbar\omega_q} - 1)$ is the Bose-Einstein distribution function, and $f(\epsilon_{k+q}) = 1/(e^{\beta\epsilon_{k+q}} + 1)$ is the Fermi-Dirac distribution function. We focus on the conduction band of Ag and start with the band structure obtained from GW calculations shown in Fig. 2(g) in the main text. For simplification, we further approximate the conduction band around Γ point of the Brillouin zone (BZ) by a parabolic band $\epsilon_k = (\hbar k)^2/2m_c$, where the effective mass $m_c = 0.94 m_0$ is obtained from fitting to GW calculation.

Because the amplitude of large-$q$ scatterings with $q > \frac{1}{d} \approx 0.154$ Å$^{-1}$ is suppressed by the static screening effect via a factor of $e^{-qd}$, the electron self-energy is dominated by contributions from small-$q$ scatterings. Around Γ point, the contribution from inter-band scatterings to self-energy is negligible due to the large direct band gap $\Delta \approx 2.45$ eV. Moreover, the conduction band bottom is ~0.7 eV above the Fermi level, which is much larger than the energy scale of room temperature, suggesting the conduction band is thermally unoccupied with $f(\epsilon_{k+q}) = 0$. In view of these facts, the real-part self-energy in Eq. (S4) reduces to

$$\Sigma_R(\omega, \boldsymbol{k}) = \sum_q |M(\boldsymbol{q})|^2 \left[ \frac{1 + N_q}{\omega - \epsilon_{k+q} - \hbar\omega_q} + \frac{N_q}{\omega - \epsilon_{k+q} + \hbar\omega_q} \right]. \quad (S5)$$

The mass renormalization is expressed as

$$\frac{m_c^*}{m_c} = \frac{1 - \partial_\omega \Sigma_R(\omega, \boldsymbol{k})}{1 + \partial_{\epsilon_k} \Sigma_R(\omega, \boldsymbol{k})} \bigg|_{\omega = \epsilon_k}, \quad (S6)$$

where $m_c^*$ denotes the renormalized effective mass of the conduction electron in Ag. For electron-phonon interaction, order of magnitude estimations of these derivations in Eq. (S6) are $\partial_\omega \Sigma_R \sim \Sigma_R/\omega_D$ and $\partial_{\epsilon_k} \Sigma_R \sim \Sigma_R/E_F$, where $\omega_D$ and $E_F$ are Debye energy of phonon and Fermi energy of the electron. In metals, $E_F$ is much larger than $\omega_D$, and mass renormalization is simplified as $Z = 1 - \partial_\omega \Sigma_R(\omega, \boldsymbol{k})|_{\omega = \epsilon_k}$. In contrast to electron-phonon interaction, here the plasmon energy spectrum is more dispersive than the conduction band of Ag, as shown in Fig. S1. Therefore, both the frequency and wavevector dependence of $\Sigma_R(\omega, \boldsymbol{k})$ should be taken into account in the calculation of mass renormalization. From Eq. (S5), we have

$$\left.\frac{\partial \Sigma_R(\omega, \boldsymbol{k})}{\partial \omega}\right|_{\omega=\epsilon_{\boldsymbol{k}}} = -\sum_{\boldsymbol{q}} |M(\boldsymbol{q})|^2 \left[\frac{1+N_q}{(\delta\epsilon - \hbar\omega_q)^2} + \frac{N_q}{(\delta\epsilon + \hbar\omega_q)^2}\right], \quad (S7)$$

and

$$\left.\frac{\partial \Sigma_R(\omega, \boldsymbol{k})}{\partial \epsilon_{\boldsymbol{k}}}\right|_{\omega=\epsilon_{\boldsymbol{k}}} = \sum_{\boldsymbol{q}} |M(\boldsymbol{q})|^2 \left[\frac{1+N_q}{(\delta\epsilon - \hbar\omega_q)^2} + \frac{N_q}{(\delta\epsilon + \hbar\omega_q)^2}\right]\left(1 + \frac{q}{k}\cos\theta\right), \quad (S8)$$

where $\delta\epsilon$ is defined as $\delta\epsilon = \epsilon_{\boldsymbol{k}} - \epsilon_{\boldsymbol{k}+\boldsymbol{q}}$ and $\theta$ denotes the angle between vectors $\boldsymbol{k}$ and $\boldsymbol{q}$. Numerical solutions of the above equations are given in Fig. S5, where we find that the derivations of $\Sigma_R$ with respect to $\omega$ and $\epsilon_{\boldsymbol{k}}$ share nearly the same magnitude while with opposite signs, resulting in very weak mass renormalization. Essentially, the plasmon energy spectrum is much more dispersive than the conduction band of Ag, as illustrated in Fig. S4. By ignoring $\delta\epsilon$ in Eqs. (S7) and (S8), we have

$$\left.\frac{\partial \Sigma_R(\omega, \boldsymbol{k})}{\partial \omega}\right|_{\omega=\epsilon_{\boldsymbol{k}}} \approx -\sum_{\boldsymbol{q}} |M(\boldsymbol{q})|^2 \frac{1+2N_q}{(\hbar\omega_q)^2}, \quad (S9)$$

and

$$\left.\frac{\partial \Sigma_R(\omega, \boldsymbol{k})}{\partial \epsilon_{\boldsymbol{k}}}\right|_{\omega=\epsilon_{\boldsymbol{k}}} \approx \sum_{\boldsymbol{q}} |M(\boldsymbol{q})|^2 \frac{1+2N_q}{(\hbar\omega_q)^2}\left(1 + \frac{q}{k}\cos\theta\right) = \sum_{\boldsymbol{q}} |M(\boldsymbol{q})|^2 \frac{1+2N_q}{(\hbar\omega_q)^2}, \quad (S10)$$

consistent with the numerical calculations. By increasing $m^*$ or decreasing the plasmon energy, the mass renormalization will be slightly enhanced as shown in Fig. S5.

Overall, the electron-plasmon interaction-induced mass renormalization of the Ag conduction band is very weak and cannot explain the factor of 2.6 enhancement observed in trARPES measurements in the main text.

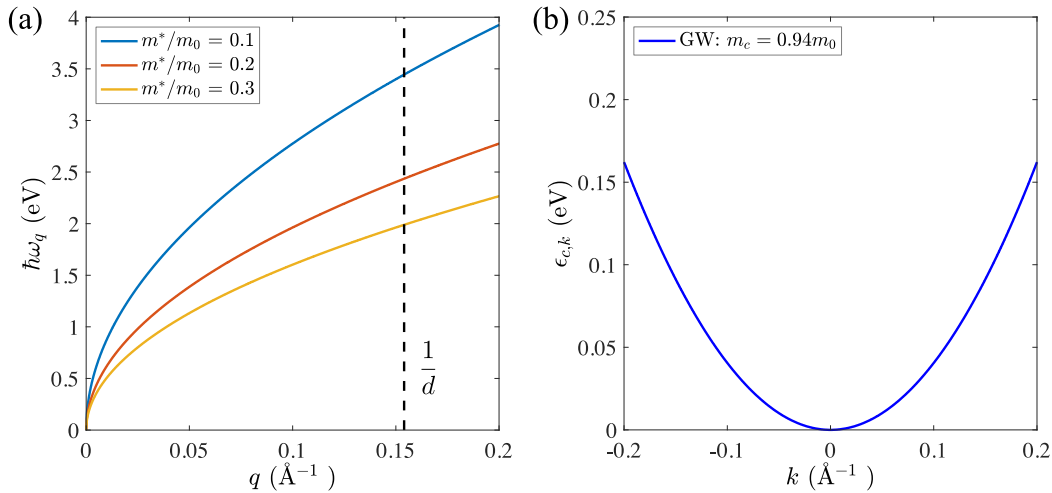

**Figure S4** (a) Plasmon dispersion of BL-Gr for different values of electron effective mass. The dashed line marks the wavevector $1/d$, above which the electron-plasmon interaction is strongly screened. (b) Parabolic band structure of the Ag conduction band was obtained by fitting GW calculations around the BZ center.

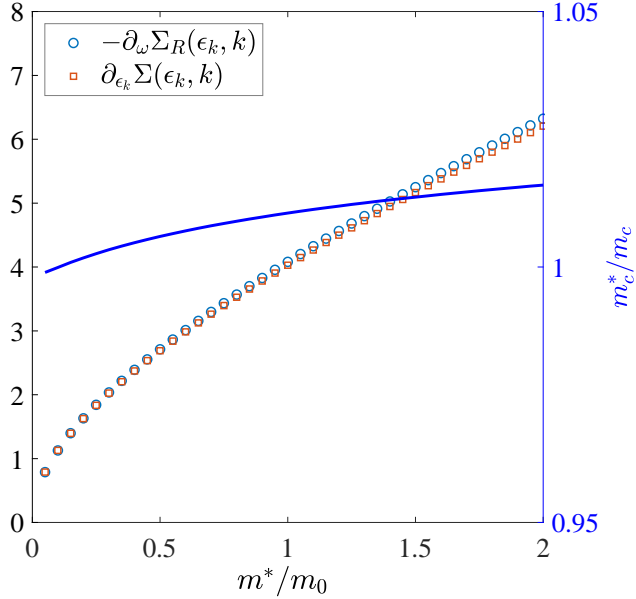

**Figure S5** Left axis: Derivation of $\Sigma_R(\omega, \boldsymbol{k})$ with respect to ω and $\epsilon_k$. Right axis: Mass renormalization of Ag conduction band upon varying the effective mass of electron $m^*$ in BL-Gr. These results are obtained by choosing $k = 0.1 \, \text{Å}^{-1}$.